\documentclass[twocolumn,showpacs,preprintnumbers,amsmath,amssymb]{revtex4}

\usepackage{graphicx}
\usepackage{dcolumn}
\usepackage{bm}

\newcommand{\bq}{\begin{equation}}
\newcommand{\eq}{\end{equation}}
\newcommand{\bqa}{\begin{eqnarray}}
\newcommand{\eqa}{\end{eqnarray}}
\newcommand{\nn}{\nonumber \\}
\newcommand{\ij}{\langle ij \rangle}

\def\be     {\begin{equation}}
\def\ee     {\end{equation}}
\def\bea        {\begin{eqnarray}}
\def\eea        {\end{eqnarray}}
\def\bnn    {\begin{eqnarray*}}
\def\enn    {\end{eqnarray*}}

\begin{document}

\title{Role of disorder in the Mott-Hubbard transition}
\author{Ki-Seok Kim}
\affiliation{School of Physics, Korea Institute for Advanced
Study, Seoul 130-012, Korea}
\date{\today}

\begin{abstract}
We investigate the role of disorder in the Mott-Hubbard transition
based on the slave-rotor representation of the Hubbard model,
where an electron is decomposed into a fermionic spinon for a spin
degree of freedom and a bosonic rotor (chargon) for a charge
degree of freedom. In the absence of disorder the Mott-Hubbard
insulator is assumed to be the spin liquid Mott insulator in terms
of gapless spinons near the Fermi surface and gapped chargons
interacting via U(1) gauge fields. We found that the Mott-Hubbard
critical point becomes unstable as soon as disorder is turned on.
As a result, a disorder critical point appears to be identified
with the spin liquid glass insulator to the Fermi liquid metal
transition, where the spin liquid glass consists of the U(1) spin
liquid and the chargon glass. We expect that glassy behaviors of
charge fluctuations can be measured by the optical spectra in the
insulating phase of an organic material
$\kappa-(BEDT-TTF)_{2}Cu_{2}(CN)_{3}$. Furthermore, since the
Mott-Anderson critical point depends on the spinon conductivity,
universality in the critical exponents may not be found.
\end{abstract}


\maketitle

\section{Introduction}

Metal-insulator transition (MIT) is one of the most studied
subjects in condensed matter physics. However, even the existence
of the MIT is not convincingly proven at zero temperature in two
spatial dimensions [$(2+1)D$], especially when both interaction
and disorder coexist.\cite{Review_MIT} The Mott-Hubbard MIT has
been claimed to occur at the critical interaction strength in the
Hubbard model without disorder.\cite{MHMIT} On the other hand, it
is believed that the Anderson MIT does not arise in two spatial
dimensions for the case of noninteracting electrons.\cite{MAMIT}
The common belief seems that the MIT in the presence of both
interaction and disorder would not appear in $(2+1)D$ because both
interaction and disorder increase an insulating tendency. Recent
experiments challenge this belief.\cite{Review_EXP} In Si
metal-oxide-semiconductor field-effect transistor, and other high
mobility semiconductor devices, unexpected MITs have been reported
although these transitions are questioned to be truly quantum
phase transitions owing to the temperature ranges in these
experiments.\cite{Das_Sarma}

In the preset paper we investigate the role of disorder in the
Mott-Hubbard MIT based on the Hubbard model. The main questions in
this paper are (1) the nature of the insulating phase and (2) the
nature of the MIT, where the nonmagnetic insulating phase is
assumed to be a spin liquid state with a Fermi surface in the
absence of disorder. The core in these questions is the fate of
gapped charge fluctuations in the Mott-Hubbard insulator when
disorder is introduced.

Using the slave-rotor representation of the Hubbard
model,\cite{SR_FG} we obtain an effective field theory for the MIT
in the presence of disorder, given by a U(1) gauge theory in terms
of fermionic spinons and bosonic collective excitations
interacting via U(1) gauge fields.\cite{SR_LeeLee,SR_Kim} Disorder
couples to charge fluctuations, and affect their dynamics
severely. Using a renormalization group (RG) analysis, we argue
that the Mott-Hubbard MIT turns into the Mott-Anderson MIT since
the pure MIT critical point becomes unstable as soon as disorder
is turned on, resulting in a new stable fixed point with a finite
disorder. Accordingly, the resulting insulating phase with
disorder is expected to be a Bose glass state, where gapped charge
excitations in the Mott-Hubbard insulator are gapless in the
presence of disorder. As a result, the U(1) spin liquid state
would coexist with the Bose glass phase of charge fluctuations,
thus called the spin liquid charge glass.

The present study is expected to apply to geometrically frustrated
lattices such as an organic material
$\kappa-(BEDT-TTF)_{2}Cu_{2}(CN)_{3}$ since the U(1) spin liquid
Mott insulator is believed to appear in this
material.\cite{SR_LeeLee} Our study implies that although spin
dynamics is little affected by weak disorder, charge dynamics is
severely modified from the Bose-Mott insulator to the Bose glass.
This will be measured by charge spectra in optical conductivity
experiments.

\section{Slave-rotor theory with disorder}

\subsection{Formulation}

We consider the Hubbard model with disorder \bqa && H = -
t\sum_{ij\sigma}c_{i\sigma}^{\dagger}c_{j\sigma} +
u\sum_{i}(\sum_{\sigma}c_{i\sigma}^{\dagger}c_{i\sigma})^{2} \nn
&& - \sum_{i}v_{i}(\sum_{\sigma}c_{i\sigma}^{\dagger}c_{i\sigma})
, \eqa where $t$ is a hopping integral, $u$ the strength of
on-site Coulomb interaction, and $v_{i}$ a random potential
introduced by disorder.

The slave-rotor representation is utilized in order to treat the
Hubbard $u$ term. Because this methodology is well introduced in
Refs. \cite{SR_FG,SR_Kim}, here we do not discuss the rotor
representation in detail. An electron annihilation operator can be
decomposed into a spin annihilation operator $f_{i\sigma}$ and a
charge one $e^{-i\theta_{i}}$ in the following way \bqa &&
c_{i\sigma} = e^{-i\theta_{i}}f_{i\sigma} . \eqa In this paper we
call $f_{i\sigma}$ and $e^{-i\theta_{i}}$ spinon and chargon,
respectively. Inserting this decomposition into Eq. (1), one can
obtain \bqa && Z =
\int{D[f_{i\sigma},\theta_{i},\varphi_{i},L_{i}]}e^{-\int{d\tau}
L} , \nn && L = \sum_{i\sigma}f_{i\sigma}^{*}(\partial_{\tau} -
\mu)f_{i\sigma} - t\sum_{ij\sigma}f_{i\sigma}^{*}e^{i(\theta_{i} -
\theta_{j})}f_{j\sigma} \nn && + \sum_{i}[uL_{i}^{2} -
iL_{i}(\partial_{\tau}\theta_{i} - iv_{i}) + i\varphi_{i}(L_{i} -
\sum_{\sigma}f_{i\sigma}^{*}f_{i\sigma})] . \nn \eqa Here
$\varphi_{i}$ is a Lagrange multiplier field imposing the rotor
constraint $L_{i} =
\sum_{\sigma}f_{i\sigma}^{\dagger}f_{i\sigma}$, and $\mu$ the
chemical potential of electrons. Physically, $\varphi_{i}$ is an
effective electric potential associated with a charge density wave
order parameter, and $L_{i}$ an electron density operator
canonically conjugate to the phase field $\theta_{i}$, indicated
by the term $-iL_{i}\partial_{\tau}\theta_{i}$. It should be noted
that Eq. (3) is just another representation of Eq. (1) via the
transformation Eq. (2). Integrating over the potential field
$\varphi_{i}$ and the density field $L_{i}$ in Eq. (3), and
performing the gauge transformation Eq. (2), one can recover the
Hubbard model Eq. (1).

Integrating out the density variable $L_{i}$, Eq. (3) reads \bqa
&& Z = \int{D[f_{i\sigma},\theta_{i},\varphi_{i}]}e^{-\int{d\tau}
L} , \nn&& L = \sum_{i\sigma}f_{i\sigma}^{*}(\partial_{\tau} - \mu
- i\varphi_{i})f_{i\sigma} -
t\sum_{ij\sigma}f_{i\sigma}^{*}e^{i(\theta_{i} -
\theta_{j})}f_{j\sigma} \nn && +
\frac{1}{4u}\sum_{i}(\partial_{\tau}\theta_{i} - \varphi_{i} -
iv_{i})^{2} . \eqa Note that the random potential $v_{i}$ couples
to the charge density represented by $\partial_{\tau}\theta_{i}$.
Decomposing the hopping term by using the Hubbard-Stratonovich
(HS) transformation, one can obtain the effective Lagrangian \bqa
&& L = L_{0} + L_{f} + L_{\theta} , \nn && L_{0} =
t\sum_{\ij}(\alpha_{ij}\beta_{ij}^{*} + \beta_{ij}\alpha_{ij}^{*})
, \nn && L_{f} = \sum_{i\sigma}f_{i\sigma}^{*}(\partial_{\tau} -
\mu - i\varphi_{i})f_{i\sigma} \nn && -
t\sum_{\ij\sigma}(f_{i\sigma}^{*}\beta_{ij}^{*}f_{j\sigma} +
f_{j\sigma}^{*}\beta_{ij}f_{i\sigma}) , \nn && L_{\theta} =
\frac{1}{4u}\sum_{i}(\partial_{\tau}\theta_{i} - \varphi_{i} -
iv_{i})^{2} \nn && -
t\sum_{\ij}(e^{i\theta_{i}}\alpha_{ij}e^{-i\theta_{j}} +
e^{i\theta_{j}}\alpha_{ij}^{*}e^{-i\theta_{i}}) , \eqa where
$\alpha_{ij}$ and $\beta_{ij}$ are spinon and chargon hopping
order parameters, respectively.

A saddle point analysis results in the self-consistent equations
\bqa && \langle\sum_{\sigma}f_{i\sigma}^{*}f_{i\sigma}\rangle = 1
, \nn && -i{\varphi} = - i\langle\partial_{\tau}\theta_{i}\rangle
+ 2u\langle\sum_{\sigma}f_{i\sigma}^{*}f_{i\sigma}\rangle = 2u ,
\nn && \alpha_{ij} =
\langle\sum_{\sigma}{f}_{i\sigma}^{*}f_{j\sigma}\rangle , ~~~~~
\beta_{ij} = \langle{e}^{i\theta_{j}}e^{-i\theta_{i}}\rangle .
\eqa Considering low energy fluctuations around this saddle point,
one can set \bqa && \alpha_{ij} = \alpha{e}^{ia_{ij}}, ~~~~~
\beta_{ij} = \beta{e}^{ia_{ij}}, ~~~~~ \varphi_{i} = {\varphi} +
a_{i\tau} , \eqa where $\alpha =
|\langle\sum_{\sigma}{f}_{i\sigma}^{*}f_{j\sigma}\rangle|$ and
$\beta = |\langle{e}^{i\theta_{j}}e^{-i\theta_{i}}\rangle|$ are
amplitudes of the hopping order parameters, and $a_{ij}$ and
$a_{i\tau}$ are spatial and time components of U(1) gauge fields.

Inserting Eq. (7) into Eq. (5), we find an effective U(1) gauge
theory for the Mott-Anderson transition \bqa && L_{f} =
\sum_{i\sigma}f_{i\sigma}^{*}(\partial_{\tau} -
ia_{i\tau})f_{i\sigma} \nn && -
t\beta\sum_{\ij\sigma}(f_{i\sigma}^{*}e^{-ia_{ij}}f_{j\sigma} +
h.c.) , \nn && L_{\theta} =
\frac{1}{4u}\sum_{i}(\partial_{\tau}\theta_{i} - a_{i\tau} -
iv_{i})^{2} \nn && - 2t\alpha\sum_{\ij}\cos(\theta_{j} -
\theta_{i} - a_{ij}) , \eqa where the mean field potential
$\varphi$ cancels the chemical potential at half filling in the
fermion Lagrangian, and the Berry phase term $S_{B} =
i\sum_{i}\int_{0}^{\beta}{d\tau}\partial_{\tau}\theta_{i}$
resulting from the mean field potential $\varphi$ has no physical
effects at half filling in the boson Lagrangian, thus safely
ignored.\cite{SR_Kim}

\subsection{Discussion in the Mean field level}

It is interesting to note that spinon dynamics is decoupled to
chargon dynamics in the mean field scheme ignoring gauge
fluctuations, governed by \bqa && L_{f} =
\sum_{i\sigma}f_{i\sigma}^{*}\partial_{\tau}f_{i\sigma} -
t\beta\sum_{\ij\sigma}(f_{i\sigma}^{*}f_{j\sigma} + h.c.) , \nn &&
L_{\theta} = \frac{1}{4u}\sum_{i}(\partial_{\tau}\theta_{i} -
iv_{i})^{2} - 2t\alpha\sum_{\ij}\cos(\theta_{j} - \theta_{i}) .
\eqa Since the chargon Lagrangian corresponds to the quantum XY
model in this level of approximation, the coherent-incoherent
transition of the phase fields occurs in the absence of disorder,
belonging to the XY universality class at zero temperature. The
incoherent phase with charge gap but no spin gap is identified
with the Mott-Hubbard insulator of a spin liquid with a Fermi
surface. In the spin liquid there is no coherent quasiparticle
peak at zero energy, and only incoherent hump is observed near the
correlation energy $\pm u/2$.\cite{SR_FG} On the other hand, the
coherent phase is understood as a Fermi liquid with a coherent
quasiparticle peak at zero energy.\cite{SR_FG}

Realizing that the quantum XY model $L_{\theta}$ in Eq. (9) is
equivalent to the boson Hubbard model for critical phenomena, one
can find its effective field theoretic action given
by\cite{Fisher_BG} \bqa && S = \int{d\tau}{d^2r} \Bigl[
|\partial_{\mu}\psi|^{2} + m_{\psi}^{2}|\psi|^{2} +
\frac{u_{\psi}}{2}|\psi|^{4} \nn && +
v\psi^{\dagger}\partial_{\tau}\psi + w|\psi|^{2} \Bigr] . \eqa
Here $\psi \sim \langle{e^{i\theta}}\rangle$ is the effective
chargon field with its mass $m_{\psi}^{2} \sim u/t - (u/t)_{c}$,
where $(u/t)_{c}$ is the critical strength of local interactions,
associated with the Mott transition in the mean field
level.\cite{SR_FG} $u_{\psi}$ is a phenomenologically introduced
parameter for local chargon interactions. $v$ is a Gaussian random
variable resulting from the Berry phase contribution, and $w$ a
Gaussian random mass originating from the random chemical
potential in the boson Hubbard model, where they satisfy
$\langle{v}(r)\rangle = 0$, $\langle{v}(r)v(r')\rangle =
V\delta(r-r')$ and $\langle{w}(r)\rangle = 0$,
$\langle{w}(r)w(r')\rangle = W\delta(r-r')$, respectively.

To take into account the random variables, one can utilize the
standard replica method, then obtain \bqa && Z = \int{D\psi_{l}}
e^{-S} , \nn && S = \sum_{l}\int{d\tau}{d^2r} \Bigl[
|\partial_{\mu}\psi_{l}|^{2} + m_{\psi}^{2}|\psi_l|^{2} +
\frac{u_{\psi}}{2}|\psi_l|^{4} \Bigr] \nn && -
\sum_{l,l'}\int{d\tau}{d\tau'} \int{d^2r}
\frac{V}{2}(\psi_{l\tau}^{\dagger}\partial_{\tau}\psi_{l\tau})(\psi_{l'\tau'}^{\dagger}\partial_{\tau'}\psi_{l'\tau'})
\nn && - \sum_{l,l'}\int{d\tau}{d\tau'}
\int{d^2r}\frac{W}{2}|\psi_{l\tau}|^{2}|\psi_{l'\tau'}|^{2} , \eqa
where $l, l' = 1, ..., N$ are replica indices, and the limit of $N
\rightarrow 0$ is performed in the final stage of calculations.

The pure Mott critical point ($m_{\psi}^{2*} = 0$, $u_{\psi}^{*}
\not= 0 $, $V^{*} = 0$, and $W^{*} = 0$) can be easily checked to
be unstable against the presence of disorder ($W \not= 0$). In Eq.
(11) the bare scaling dimension of $\psi$ is given by $[\psi] =
L^{1/2}$ owing to the quadratic derivative in the kinetic energy
term, where $L$ is a length scale. Thus, we obtain $[W] = L^{2}$,
indicating that disorder is relevant at the pure Mott critical
point. $[V] = L^{0}$ is obtained, thus marginal. In the
$1/N_{\psi}$ approximation where $N_{\psi}$ is the flavor number
of boson fields, it was shown that $V$ is irrelevant, and $W$
relevant so that the RG flow goes to the strong disorder
regime.\cite{YB_BG} But, recent RG calculations exhibit a weak
disorder fixed point in the case of $V = 0$, identified with the
Bose glass to superfluid transition instead of the Bose-Mott
insulator to superfluid
transition.\cite{Weak_disorder_fixed_point} The weak disorder
fixed point can be also shown to exist in the dual vortex
formulation, where the quantum XY model is mapped into the scalar
quantum electrodynamics in $(2+1)D$ (QED$_3$) in terms of vortices
interacting via vortex gauge fields. A random mass term for the
vortices is also induced by disorder, making it unstable the pure
Mott critical point associated with the Bose-Mott insulator to
superfluid transition. A new stable disorder fixed point appears
to be identified with the Bose glass to superfluid transition in
the dual formulation.\cite{Herbut_BG} In this respect the spin
liquid Mott insulator turns into the spin liquid Bose glass in the
mean field level.

Beyond the mean field level, dynamics of spinons and chargons is
coupled via U(1) gauge fluctuations. In this case several
important issues arise even in the absence of disorder. One can
doubt the stability of the spin liquid phase against U(1) gauge
fluctuations, especially instanton excitations allowed by the
compactness of the U(1) gauge field. The present author discussed
the confinement-deconfinement problem in the presence of the Fermi
surface, and proposed the stability of the spin liquid phase when
the spinon conductivity is sufficiently
large.\cite{Kim_Deconfine_FS} In addition, the XY transition
nature in the mean field level without disorder should be modified
by spinon excitations. Especially, gapless spinon excitations
result in dissipative dynamics of the U(1) gauge field. These
damped gauge fluctuations are expected to turn the XY transition
into the other. Furthermore, the presence of disorder makes the
MIT much more complex.

\section{Renormalization group analysis}

\subsection{Boson-only effective action}

To investigate the role of spinon excitations in the
coherent-incoherent transition of chargon fields, we obtain the
effective chargon-gauge action in the continuum limit by
integrating out spinons in Eq. (8) \bqa && S_{eff} =
\int{d\tau}{d^2r} \Bigl[ \frac{1}{4u}(\partial_{\tau}\theta -
a_{\tau} - iv)^{2} - 2t\alpha\cos({\nabla}\theta - \mathbf{a})
\Bigr] \nn && +
\frac{1}{\beta}\sum_{\omega_{n}}\int{dq_{r}}\frac{1}{2}a_{\mu}(q_r,i\omega_{n})
D_{\mu\nu}^{-1}(q_r,i\omega_{n})a_{\nu}(-q_r,-i\omega_{n}) , \nn
\eqa where $D_{\mu\nu}(q_r,i\omega_{n})$ is the renormalized gauge
propagator, given by \bqa && D_{\mu\nu}(q_r,i\omega_{n}) =
\Bigl(\delta_{\mu\nu} -
\frac{q_{\mu}q_{\nu}}{q^2}\Bigr)D(q_r,i\omega_{n}) , \nn &&
D^{-1}(q_{r},i\omega_{n}) = D_{0}^{-1}(q_{r},i\omega_{n}) +
\Pi(q_r,i\omega_{n}) . \eqa Here $D_{0}^{-1}(q_{r},i\omega_{n}) =
(q_r^2 + \omega_{n}^2)/g^{2}$ is the bare gauge propagator given
by the Maxwell gauge action, resulting from integration of high
energy fluctuations of spinons and chargons. $g$ is an internal
gauge charge of the spinon and chargon. $\Pi(q_r,i\omega_{n})$ is
the self-energy of the gauge field, given by the correlation
function of spinon charge (number) currents. Since the
current-current correlation function is calculated in the
noninteracting fermion ensemble, its structure is well
known\cite{Lee_Nagaosa,Tsvelik} \bqa && \Pi(q_r,i\omega_{n}) =
\sigma(q_{r})|\omega_{n}| + \chi{q}_{r}^{2} . \eqa Here the spinon
conductivity $\sigma(q_{r})$ is given by $\sigma(q_{r}) \approx
k_{0}/q_{r}$ in the clean limit while it is $\sigma(q_{r}) \approx
\sigma_{0} = k_{0}l$ in the dirty limit, where $k_{0}$ is of order
$k_{F}$ (Fermi momentum), and $l$ the spinon mean free path
determined by disorder scattering. The diamagnetic susceptibility
$\chi$ is given by $\chi \sim m_{f}^{-1}$, where $m_{f} \sim
(t\beta)^{-1}$ is the band mass of spinons. The frequency part of
the kernel $\Pi(q,i\omega_n)$ shows the dissipative propagation of
the gauge field owing to particle-hole excitations of spinons near
the Fermi surface.

Recently, the present author investigated the Mott-Hubbard MIT
based on the effective chargon-gauge action Eq. (12) without
disorder.\cite{SR_Kim} In this study we found that dissipative
gauge fluctuations result in a new critical point, depending on
the spinon conductivity $\sigma_{0}$ that determines the strength
of dissipation. In the limit of $\sigma_{0} \rightarrow \infty$
identified with a perfect metal of spinons, gauge fluctuations are
completely screened by spinon excitations, thus safely ignored.
The resulting chargon action is nothing but the XY Lagrangian,
yielding the XY transition. On the other hand, in the limit of
$\sigma_{0} \rightarrow 0$ considered as an insulator of spinons,
only the Maxwell gauge action is expected to appear from high
energy contributions of spinons and chargons. The resulting
chargon-gauge action coincides with the scalar QED$_3$, yielding
the inverted XY (IXY) transition\cite{Kleinert} owing to gauge
excitations. Varying the spinon conductivity, these two limits
would be connected.

Emergence of the new charged fixed point can be easily understood
from the effective gauge-only action at the critical point.
Integrating over critical chargon fluctuations in Eq. (12), the
critical gauge action can be obtained in a highly schematic form
at the critical point \bqa && {S}_{g} =
\frac{1}{\beta}\sum_{\omega_{n}}\int{d^{2}q_{r}}
\frac{1}{2}\mathbf{a}^{T}(q_{r},i\omega_{n})\Pi(q_{r},i\omega_{n})\mathbf{a}^{T}(-q_{r},-i\omega_{n})
, \nonumber \eqa where $\mathbf{a}^{T}(q_{r},i\omega_{n})$
represent the transverse components of the gauge fields. The gauge
kernel $\Pi(q_{r},i\omega_{n})$ is given by \bqa &&
\Pi(q_{r},i\omega_{n}) =
\frac{N_{\theta}}{8}\sqrt{q_{r}^{2}+\omega_{n}^{2}} +
\sigma_{0}|\omega_{n}| , \nonumber \eqa where $N_{\theta}$ is the
flavor number of the chargon field, here $N_{\theta} = 1$. The
first term results from critical chargon fluctuations while the
second originates from gapless spinon excitations near the Fermi
surface. This gauge action can be easily checked to be
scale-invariant at the tree level, giving the IXY fixed point in
the $\sigma_{0} \rightarrow 0$ limit and the XY one in the
$\sigma_{0} \rightarrow \infty$ limit. Thus, the spinon
contribution characterized by the spinon conductivity $\sigma_{0}$
connects these two fixed points smoothly. A finite conductivity
causes a new critical point between the XY and IXY fixed points.
In the present paper we examine the role of disorder in the new
fixed point.

Before closing this section, we summarize the effective boson-only
action depending on the spinon conductivity $\sigma_{0}$ in Table
I.
\begin{table*}
\caption{Effective action depending the spinon conductivity}
\begin{tabular}{cccccccc}
\hline & $\sigma_{0} \rightarrow \infty$ & $0 < \sigma_{0} <
\infty$ & $\sigma_{0} \rightarrow 0$ \nn & Spinon perfect metal &
Spinon metal & Spinon insulator \nn \hline Effective chargon
action & XY & QED$_3$ $+$ spinon-gauge correction & QED$_3$ \nn
Dual vortex action & QED$_3$ & QED$_3$ $+$ spinon-gauge correction
& XY \nn \hline
\end{tabular}
\end{table*}

\subsection{Dual vortex action with disorder}

Disorder effects produce random Berry phase to chargon fields.
Because the Berry phase term leads to a complex phase factor to
the partition function of Eq. (12), it is not easy to handle the
partition function in the chargon representation. Duality
transformation is generally performed to treat the Berry phase
term.\cite{Fisher_Lee} The dual vortex action of Eq. (12) is
obtained to be \bqa && S_{v} = \int{d\tau}{d^2r}
\Bigl[|(\partial_{\mu} - ic_{\mu})\Phi|^{2} + m_{v}^{2}|\Phi|^{2}
+ \frac{u_{v}}{2}|\Phi|^{4} \nn && +
u(\partial\times{c})_{\tau}^{2} +
\frac{1}{4t\alpha}(\partial\times{c})_{r}^{2} -
v(\partial\times{c})_{\tau} - ia_{\mu}(\partial\times{c})_{\mu}
\Bigr] \nn && +
\frac{1}{\beta}\sum_{\omega_{n}}\int{dq_{r}}\frac{1}{2}a_{\mu}(q_r,i\omega_{n})
D_{\mu\nu}^{-1}(q_r,i\omega_{n})a_{\nu}(-q_r,-i\omega_{n}) . \nn
\eqa Here $\Phi$ is a vortex field, and $c_{\mu}$ a vortex gauge
field. $m_{v}$ is a vortex mass, given by $m_{v}^{2} \sim
(u/t)_{c} - u/t$ with the mean field MIT critical point
$(u/t)_{c}$, and $u_{v}$ a phenomenologically introduced parameter
for local interactions between vortices. The random potential $v$
plays the role of random magnetic fields in vortices.

Since Eq. (15) is quadratic in gauge fluctuations $a_{\mu}$, one
finds the effective vortex-gauge action by performing the Gaussian
integration for the gauge fields $a_{\mu}$ \bqa && Z_{v} =
\int{D[\Phi,c_{\mu}]} e^{- S_{v}} , \nn && S_{v} =
\int{d\tau}{d^2r} \Bigl[ |(\partial_{\mu} - ic_{\mu})\Phi|^{2} +
m_{v}^{2}|\Phi|^{2} + \frac{u_{v}}{2}|\Phi|^{4} \nn && +
u(\partial\times{c})_{\tau}^{2} +
\frac{1}{4t\alpha}(\partial\times{c})_{r}^{2} -
v(\partial\times{c})_{\tau}\Bigr] \nn && +
\int{d\tau}{d\tau_1}d^2rd^2r_1\frac{1}{2}c_{\mu}(r,\tau)K_{\mu\nu}(r-r_1,\tau-\tau_1)c_{\nu}(r_1,\tau_1)
, \nn \eqa where the renormalized gauge propagator
$K_{\mu\nu}(r-r_1,\tau-\tau_1)$ is given by in energy-momentum
space \bqa && K_{\mu\nu}(q_{r},i\omega_{n}) =
\Bigl(\delta_{\mu\nu} -
\frac{q_{\mu}q_{\nu}}{q^2}\Bigr)K(q_{r},i\omega_{n}) , \nn &&
K(q_r,i\omega_n) = \frac{q_{r}^2 + \omega_{n}^{2}}{(q_r^2 +
\omega_{n}^2)/g^{2} + \sigma(q_{r})|\omega_{n}| + \chi{q}_{r}^{2}}
\nn && \approx \frac{q_{r}^2 + \omega_{n}^{2}}{(q_r^2 +
\omega_{n}^2)/\overline{g}^{2} + \sigma(q_{r})|\omega_{n}|} . \eqa
Here $\overline{g}$ is a redefined variable including the
susceptibility. In the following we consider dirty cases
characterized by $\sigma(q_{r}) = \sigma_{0}$.

In order to take into account the random potential by disorder, we
use the replica trick to average over disorder. The random
magnetic field $v$ in the vortex action Eq. (16) would cause \bqa
- \sum_{l,l'}\int{d\tau}{d\tau_1}\int{d^2r}\frac{\Im}{2}
(\partial\times{c}_{l})_{\tau}(\partial\times{c}_{l'})_{\tau_1}
\nonumber \eqa for the Gaussian random potential satisfying
$\langle{v}(r)\rangle = 0$ and $\langle{v}(r)v(r_1)\rangle = \Im
\delta(r-r_1)$ with the strength $\Im$ of the random potential.
However, inclusion of only this correlation term is argued to be
not enough for disorder effects. Because the gauge-field
propagator has off-diagonal components in replica indices, the
vortex-gauge interaction of the order $\Im^{2}e_{v}^{4}$ generates
a quartic term including the couplings of different replicas of
vortices even if this term is absent initially.\cite{Herbut_BG}
Here $e_{v}$ is a vortex charge. The resulting disordered vortex
action is obtained to be \bqa && Z_{v} =
\int{D[\Phi_{l},c_{\mu{l}}]} e^{- S_{v}} , \nn && S_{v} =
\sum_{l}\int{d\tau}{d^2r} \Bigl[ |(\partial_{\mu} -
ic_{\mu{l}})\Phi_{l}|^{2} + m_{v}^{2}|\Phi_{l}|^{2} +
\frac{u_{v}}{2}|\Phi_{l}|^{4} \nn && +
u(\partial\times{c}_{l})_{\tau}^{2} +
\frac{1}{4t\alpha}(\partial\times{c}_{l})_{r}^{2} \Bigr] \nn && +
\sum_{l}\sum_{\omega_{n}}\int{dq_{r}}\frac{1}{2}c_{\mu{l}}(q_r,i\omega_n)K_{\mu\nu}(q_r,i\omega_{n})c_{\nu{l}}(-q_r,-i\omega_{n})
\nn && -
\sum_{l,l'}\int{d\tau}{d\tau_1}\int{d^2r}\frac{W}{2}|\Phi_{l\tau}|^{2}|\Phi_{l'\tau_1}|^{2}
\nn && -
\sum_{l,l'}\int{d\tau}{d\tau_1}\int{d^2r}\frac{\Im}{2}(\partial\times{c}_{l})_{\tau}(\partial\times{c}_{l'})_{\tau_1}
\eqa with $W > 0$. The correlation term induced by disorder \bqa -
\sum_{l,l'}\int{d\tau}{d\tau_1}\int{d^2r}\frac{W}{2}|\Phi_{l\tau}|^{2}|\Phi_{l'\tau_1}|^{2}
\nonumber \eqa has the same form with the term resulting from a
random mass term. Eq. (18) is our starting action for studying the
role of disorder in the Mott-Hubbard MIT.

\subsection{Renormalization group analysis}

We perform an RG analysis for Eq. (18). Anisotropy in the Maxwell
gauge action for the vortex gauge field is assumed to be
irrelevant, and only the isotropic Maxwell gauge action is
considered by replacing $u, 1/4t\alpha$ with $1/(2e_{v}^{2})$,
where $e_{v}$ is a vortex charge. In the limit of small anisotropy
the anisotropy was shown to be irrelevant at one loop
level.\cite{Herbut_BG} Furthermore, the correlation term between
random magnetic fluxes is also ignored. In the small $\Im$ limit
this term was shown to be exactly marginal at one loop
level.\cite{Herbut_BG} To address the quantum critical behavior at
the Mott transition, we introduce the scaling $r = e^{l}r'$ and
$\tau = e^{l}\tau'$, and consider the renormalized theory at the
transition point $m_{v}^{2} = 0$ \bqa && S_{v} =
\sum_{l}\int{d\tau'}{d^{2}r'} \Bigl[ Z_{\Phi}|(\partial'_{\mu} -
ie_{v}c_{\mu{l}})\Phi_{l}|^{2} \nn && +
Z_{u}\frac{u_{v}}{2}|\Phi_{l}|^{4} +
\frac{Z_{c}}{2}(\partial'\times{c}_{l})^{2} \Bigr] \nn && -
\sum_{l,l'}\int{d\tau'}{d\tau'_1}\int{d^{2}r'}Z_{W}\frac{W}{2}|\Phi_{l\tau}|^{2}|\Phi_{l'\tau_1}|^{2}
, \eqa where $Z_{\Phi}$, $Z_{u}$, $Z_{c}$, and $Z_{W}$ are the
renormalization factors defined by \bqa && \Phi =
e^{-\frac{1}{2}l}Z_{\Phi}^{\frac{1}{2}}\Phi_{r} \mbox{, } \mbox{ }
c_{\mu} = e^{-\frac{1}{2}l}Z_{c}^{\frac{1}{2}}c_{\mu{r}} , \nn &&
e_{v}^2 = e^{-l}Z_{c}^{-1}e_{vr}^{2} \mbox{,    } \mbox{ } u_{v} =
e^{-l}Z_{u}Z^{-2}_{\Phi}u_{vr} , \nn && W =
e^{-2l}Z_{W}Z^{-2}_{\Phi}W_{r} . \eqa In the renormalized action
Eq. (19) the subscript $r$ implying "renormalized" is omitted for
simple notation.

Evaluating the renormalization factors at one loop level, the RG
equations are expected to
be\cite{Weak_disorder_fixed_point,Herbut_BG,Kleinert,Ye,Derivation}
\bqa && \frac{de_{v}^{2}}{dl} = e_{v}^{2} - \Bigl(\lambda +
\frac{\zeta}{\sigma_{0}} \Bigr){e}_{v}^{4} , \nn &&
\frac{du_{v}}{dl} = u_{v} + \Bigl(h(\sigma_{0},e_{v}^{2})e_{v}^{2}
+ \gamma{W} \Bigr)u_{v} - \rho{u}_{v}^{2} -
g(\sigma_{0},e_{v}^{2})e_{v}^{4} , \nn && \frac{dW}{dl} = 2W +
\Bigl(h(\sigma_{0},e_{v}^{2})e_{v}^{2} - \kappa{u}_{v} \Bigr)W +
\eta{W}^{2} . \eqa Here $\lambda,\zeta,\gamma,\rho,\kappa,\eta$
are positive numerical constants, and
$h(\sigma_{0},e_{v}^{2}),g(\sigma_{0},e_{v}^{2})$ are analytic and
monotonically increasing functions of $\sigma_{0}$, as will be
explained below.

The first RG equation for the vortex charge can be understood in
the following way. Integrating out critical vortex fluctuations,
we obtain the singular contribution for the effective gauge action
\bqa && S_{c} = \frac{1}{\beta}\sum_{\omega_{n}}\int{d^{2}q_{r}}
\frac{1}{2}
c_{\mu}(q_r,i\omega_n)\Xi_{\mu\nu}(q_r,i\omega_{n})c_{\nu}(-q,-i\omega_{n})
, \nn && \Xi_{\mu\nu}(q_r,i\omega_{n}) = \Bigl(\delta_{\mu\nu} -
\frac{q_{\mu}q_{\nu}}{q^2}\Bigr)\Xi(q_r,i\omega_{n}) , \nn &&
\Xi(q_r,i\omega_{n}) = \frac{N_{v}}{8}\sqrt{q_{r}^2 +
\omega_{n}^{2}} + K(q_{r},i\omega_{n}) \nn && \approx
\frac{N_{v}}{8}\sqrt{q_{r}^2 + \omega_{n}^{2}} + \frac{q_{r}^2 +
\omega_{n}^{2}}{\sigma_{0}|\omega_{n}|} , \nonumber \eqa where
$N_{v}$ is the flavor number of the vortex field, here $N_{v} =
1$. The first term in the kernel $\Xi(q_r,i\omega_{n})$ results
from the screening effect of the vortex charge via vortex
polarization, causing the $-\lambda{e}_{v}^{4}$ term in the RG
equation while the second originates from that via spinon
excitations, yielding the $- ({\zeta}/{\sigma_{0}})e_{v}^{4}$
term. The first $e_{v}^{2}$ term in the RG equation denotes the
bare scaling dimension of the vortex charge in $(2+1)D$. We note
that the above critical gauge action results in the relativistic
dispersion $\omega \sim q_{r}$.

For the second and third RG equations, unfortunately, we do not
know the exact functional forms of $h(\sigma_{0},e_{v}^{2})$ and
$g(\sigma_{0},e_{v}^{2})$ owing to the complexity of the gauge
kernel. Owing to the spinon contribution $K(q_{r},i\omega_{n})$
[Eq. (17)] the kernel of the gauge propagator ($c_{\mu}$) \bqa &&
D_{c}(q_{r},i\omega_{n}) = \frac{1}{q_{r}^{2} + \omega_{n}^{2} +
e_{v}^{2}K(q_{r},i\omega_{n})} \nn && \approx
\frac{\sigma_{0}|\omega_{n}|}{(q_{r}^{2} +
\omega_{n}^{2})(e_{v}^{2} + \sigma_{0}|\omega_{n}|)} \nonumber
\eqa should be utilized instead of the Maxwell propagator in
calculating one loop diagrams. Note the dependence of the vortex
charge $e_{v}^{2}$ in the effective gauge propagator. This gives
the dependence of the vortex charge to the analytic functions
$h(\sigma_{0},e_{v}^{2})$ and $g(\sigma_{0},e_{v}^{2})$. Although
the exact functional forms are not known, the limiting values of
these functions can be found.

\subsubsection{$\sigma_{0} \rightarrow \infty$}

In the limit of $\sigma_{0} \rightarrow \infty$ the gauge kernel
is reduced to the Maxwell propagator \bqa D_{c}(q_{r},i\omega_{n})
= \frac{1}{(q_{r}^{2} + \omega_{n}^{2})} \nonumber \eqa because
gauge fluctuations $a_{\mu}$ are completely screened via spinon
excitations in the perfect spinon metal, ignored and the resulting
chargon action is nothing but the quantum XY model, causing the
scalar QED$_3$ as an effective vortex-gauge action. See Table I.
Thus, $h(\sigma_{0} \rightarrow \infty,e_{v}^{2}) \rightarrow
c_{1}$ and $g(\sigma_{0} \rightarrow \infty,e_{v}^{2}) \rightarrow
c_{2}$ are obtained, where $c_{1}$ and $c_{2}$ are positive
numerical constants. Then, Eqs. (21) become the RG equations of
the scalar QED$_3$ with a random mass term \bqa &&
\frac{de_{v}^{2}}{dl} = e_{v}^{2} - \lambda{e}_{v}^{4} , \nn &&
\frac{du_{v}}{dl} = (1 + c_{1}e_{v}^{2} + \gamma{W})u_{v} -
\rho{u}_{v}^{2} - c_{2}e_{v}^{4} , \nn && \frac{dW}{dl} = (2 +
c_{1}e_{v}^{2} - \kappa{u}_{v})W + \eta{W}^{2} . \eqa These are
formally the same as the RG equations studied in Ref.
\cite{Herbut_BG}, where the existence of the weak disorder fixed
point was nicely discussed, guaranteeing the presence of the
Mott-Anderson MIT.

In the absence of disorder ($W^{*} = 0$) a stable charged critical
point ($e_{v}^{*2} \not= 0$) is expected to appear, associated
with the Mott insulator to superfluid transition although there is
a delicate issue about the existence of the charged fixed point
when the flavor number of complex matter fields is one,
corresponding to the superconducting transition. This issue is
well discussed in Ref. \cite{Kleinert}. In this paper we assume
the existence of the charged Mott critical point. This fixed point
becomes unstable as soon as disorder is turned on, as shown in the
third RG equation for $W$. A new stable fixed point is found with
a finite disorder ($W^{*} \not= 0$), identified with the Bose
glass to superfluid critical point.\cite{Herbut_BG}

\subsubsection{$\sigma_{0} \rightarrow 0$}

In the spinon insulator of $\sigma_{0} \rightarrow 0$ the
chargon-gauge action is given by the scalar QED$_3$ with disorder,
as discussed before. The resulting vortex action becomes the
$\Phi^{4}$ model with a random mass term since vortex gauge
fluctuations $c_{\mu}$ are gapped owing to the presence of long
range interactions mediated by the U(1) gauge fields $a_{\mu}$,
thus ignored in the low energy limit. This coincides with the fact
that the gauge kernel $D_{c}(q_{r},i\omega)$ vanishes. As a
result, $h(\sigma_{0} \rightarrow 0,e_{v}^{2}) \rightarrow 0$ and
$g(\sigma_{0} \rightarrow 0,e_{v}^{2}) \rightarrow 0$ are
obtained. Accordingly, Eqs. (21) are reduced to the RG equations
of the $\Phi^{4}$ theory with a random mass term \bqa &&
\frac{du_{v}}{dl} = (1 + \gamma{W})u_{v} - \rho{u}_{v}^{2} , \nn
&& \frac{dW}{dl} = (2 - \kappa{u}_{v})W + \eta{W}^{2} . \eqa

The existence of the weak disorder fixed point can be shown in Eq.
(23)\cite{Weak_disorder_fixed_point,RG_fixed_point} although there
are some papers claiming that there is no weak disorder fixed
point in this model.\cite{Fisher_BG,YB_BG} If only the two
parameters corresponding to $u_{v}$ and $W$ are considered in the
RG equations of Ref. \cite{Herbut_BG}, one finds that there indeed
exists the weak disorder fixed point, describing the Mott-Anderson
transition. One can perform the RG analysis based on the
chargon-gauge QED$_3$ action instead of the vortex $\Phi^{4}$
action. The chargon QED$_{3}$ with a random mass term is formally
equivalent to the vortex QED$_{3}$ with a random mass term if we
correspond chargons, chargon gauge fields, and chargon random mass
to vortices, vortex gauge fields, and vortex random mass.
According to the previous discussion in the vortex QED$_3$, a
disorder critical point would be found in the chargon QED$_3$,
implying that the Bose glass to superfluid transition also appears
in this case.

\subsubsection{$0 < \sigma_{0} < \infty$}

In the small $\sigma_{0}$ limit ($\sigma_{0}|\omega_{n}| <<
e_{v}^{2}$) the gauge kernel is given by \bqa &&
D_{c}(q_{r},i\omega_{n}) \approx
\frac{\sigma_{0}}{e_{v}^{2}}\frac{|\omega_{n}|}{q_{r}^{2} +
\omega_{n}^{2}} , \nonumber \eqa thus resulting in
$h(\sigma_{0},e_{v}^{2}) = c_{h}\sigma_{0}/e_{v}^{2}$ and
$g(\sigma_{0},e_{v}^{2}) = c_{g}\sigma_{0}^{2}/e_{v}^{4}$, where
$c_{h}$ and $c_{g}$ are positive numerical constants. The
corresponding RG equations are obtained to be \bqa &&
\frac{de_{v}^{2}}{dl} = e_{v}^{2} - \Bigl(\lambda +
\frac{\zeta}{\sigma_{0}} \Bigr){e}_{v}^{4} , \nn &&
\frac{du_{v}}{dl} =(1 + c_{h}\sigma_{0} + \gamma{W})u_{v} -
\rho{u}_{v}^{2} - c_{g}\sigma_{0}^{2} , \nn && \frac{dW}{dl} = (2
+ c_{h}\sigma_{0} - \kappa{u}_{v})W + \eta{W}^{2} , \eqa where the
RG flows of $e_{v}^{2}$ and $u_{v}, W$ are decoupled in this
limit. If $\sigma_{0}$ is replaced with $e_{v}^{2}$ in the
$\sigma_{0} \rightarrow \infty$ limit, Eq. (24) coincides with Eq.
(22). The replacement of $\sigma_{0}$ with $e_{v}^{2}$ is
justified by the fact that the above gauge kernel should be
reduced to that in the $\sigma_{0} \rightarrow \infty$ limit. In
this respect Eq. (24) can be considered to be a bridge between Eq.
(22) and Eq. (23).

Ignoring the $\sigma_{0}^{2}$ term in the second RG equation, one
finds the weak disorder fixed point depending on the spinon
conductivity.\cite{RG_fixed_point2} This fixed point coincides
with that of Eq. (23) in the $\sigma_{0} \rightarrow 0$ limit. As
increasing $\sigma_{0}$, we expect that the fixed point of Eq.
(24) gets close to that of Eq. (22) because Eq. (24) should
correspond to Eq. (22), as discussed above. In other words, the
Mott-Anderson critical point is expected to move from the disorder
fixed point of the $\sigma_{0} \rightarrow 0$ limit to that of the
$\sigma_{0} \rightarrow \infty$ limit, depending on the spinon
conductivity. This can be understood in the following way. The
pure Mott critical points between the $\sigma_{0} \rightarrow
\infty$ and $\sigma_{0} \rightarrow 0$ limits are smoothly
connected by controlling the spinon conductivity, as discussed
before. The presence of disorder makes the pure Mott critical
points unstable, resulting in new disorder fixed points. Thus, it
is natural that these new disorder fixed points are also connected
smoothly through varying the spinon conductivity, as clearly shown
in the small\cite{RG_fixed_point2} and large $\sigma_{0}$ limits.
Since the critical points depend on the spinon conductivity, the
concept of universality is not applied to the Mott-Anderson
transition from the spin liquid charge glass to the Fermi liquid
metal.

\subsection{Phase diagram and discussion}

We summarize our results in the schematic phase diagram Fig. 1,
where SLBG is the spin liquid Bose glass, SLMI the spin liquid
Mott insulator, FL the Fermi liquid metal, and AI the Anderson
insulator. It should be noted that our approach cannot cover the
whole range of the phase diagram. The regions indicated by
question marks in Fig. 1 are beyond the scope of this theory.
Strictly speaking, although the slave-rotor theory can produce
meaningful physics in the Fermi liquid regime ($u/t <
(u/t)_{c}$),\cite{SR_FG} the RG equations in this paper would not
be applied because chargon condensation
$\langle{e^{i\theta_{i}}}\rangle \not= 0$ allows only electron
excitations owing to confinement between condensed chargons and
spinons. When the strength of disorder becomes large, the RG
equations would not work because the present analysis is based on
the perturbation theory for weak disorder. Furthermore, strong
disorder decreases the spinon conductivity, making the spin liquid
phase unstable against instanton excitations, as discussed
before.\cite{Kim_Deconfine_FS} Remember that the spin liquid state
can be stable in the sufficiently good spinon metal. Thus, our RG
analysis can be applied to a limited range of the phase diagram
near the MIT in the presence of weak disorder, marked by dotted
arrow lines. In the clean limit ($W \rightarrow 0$) the
Mott-Hubbard MIT is obtained between SLMI and FL.\cite{SR_FG} On
the other hand, in the small disorder limit the Mott-Hubbard MIT
is shown to turn into the Mott-Anderson MIT between SLBG and FL.
The chargon Mott insulator is expected to evolve into the chargon
Bose glass as soon as disorder is turned on, as discussed earlier.
Since the chargon superfluidity appears in the presence of
disorder, the resulting electronic phase may be identified with
the Fermi liquid metal. We note that in the weak interaction limit
$u/t << (u/t)_{c}$ the Anderson transition from FL to AI is
expected to occur by increasing disorder although this is beyond
the scope of the slave-rotor theory.

\begin{figure}
\includegraphics[width=8cm]{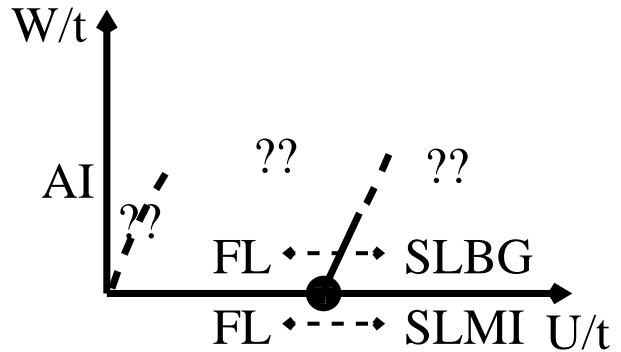}
\caption{\label{Fig. 1} A schematic phase diagram in the
slave-rotor representation of the Hubbard model with disorder}
\end{figure}

A recent dynamical mean field theory (DMFT) study shows that the
nonmagnetic phase with weak disorder is still a Mott
insulator.\cite{DMFT} The Mott insulating phase in the DMFT study
seems to be in contrast to our claim that the paramagnetic phase
is a gapless insulator of the Anderson type instead of the Mott
one. We argue that this difference is not a contradiction because
physics of the DMFT approach differs from that of our approach.
The paramagnetic Mott insulator in the DMFT study is different
from the spin liquid Mott insulator in the slave-rotor theory in
that (1) elementary spin excitations carry the spin quantum number
$1$ instead of the fractionalized spin $1/2$, and (2) there is no
spin-charge separation physics. As mentioned in the introduction,
the spin liquid Mott insulator is possible to appear in the
triangular lattice such as an organic material
$\kappa-(BEDT-TTF)_{2}Cu_{2}(CN)_{3}$.\cite{SR_LeeLee} The present
slave-rotor theory is expected to apply to the triangular lattice
while the DMFT study would explain the square lattice. In this
respect these two approaches see different systems, thus the
resulting disordered insulators can be different.

We should point out an important issue that the nature of the
insulating phase in the boson Hubbard model with weak disorder is
not completely understood. The Bose-Mott insulator was claimed at
commensurate filling instead of the Bose glass
insulator.\cite{BM_BG} One different thing from our vortex
formulation is that the present vortex action includes anomalous
gauge interactions resulting from the spinon contribution.

\section{Fermion-only effective theory with disorder}

There is an alternative way treating disorder in the slave-rotor
representation of the Hubbard model. In the effective gauge
Lagrangian Eq. (8) the gauge shift $a_{i\tau} \rightarrow
a_{i\tau} - iv_{i}$ results in \bqa && L_{f} =
\sum_{i\sigma}f_{i\sigma}^{*}(\partial_{\tau} -
ia_{i\tau})f_{i\sigma} -
\sum_{i\sigma}v_{i}f_{i\sigma}^{\dagger}f_{i\sigma} \nn && -
t\beta\sum_{\ij\sigma}(f_{i\sigma}^{*}e^{-ia_{ij}}f_{j\sigma} +
h.c.) , \nn && L_{\theta} =
\frac{1}{4u}\sum_{i}(\partial_{\tau}\theta_{i} - a_{i\tau})^{2} -
2t\alpha\sum_{\ij}\cos(\theta_{j} - \theta_{i} - a_{ij}) . \nn
\eqa Interestingly, the effect of disorder appears as the random
chemical potential of spinons instead of the random Berry phase of
chargons. Since the chargon dynamics does not couple to disorder
directly, the glass phase is not likely to appear in this
approach. Only the Mott insulator to superfluid transition is
expected to occur in the chargon Lagrangian.

Integrating out the gapped chargon excitations in the Mott
insulating phase, we obtain an effective spinon-gauge action in
the continuum limit \bqa && S_{f} = \int{d\tau}{d^2r} \Bigl[
\sum_{\sigma}\Bigl( f_{\sigma}^{\dagger}(\partial_{\tau} -
ia_{\tau})f_{\sigma} + \frac{1}{2m_{f}}|(\partial_{r} -
ia_{r})f_{\sigma}|^{2} \nn && - vf_{\sigma}^{\dagger}f_{\sigma}
\Bigr) + \frac{1}{2g^2}|\partial\times{a}|^{2} \Bigr] , \eqa where
$m_{f} \sim (t\beta)^{-1}$ is a spinon band mass. The main
question in this effective spinon action is about the role of
random chemical potentials in the spinon dynamics.

The role of nonmagnetic disorder in the QED$_3$ without the Fermi
surface was investigated by the present
author.\cite{Kim_disorder,Kim_disorder_sigma} In contrast to the
$(2+1)D$ free Dirac theory long range gauge interactions are shown
to reduce the strength of disorder, and induce a delocalized state
in the QED$_3$. The presence of disorder destabilizes the free
Dirac fixed point. The RG flow goes away from the fixed point,
indicating localization.\cite{Lee_disorder,Fisher_disorder} On the
other hand, the charged fixed point in the QED$_3$ remains stable
at least against weak randomness. A new unstable fixed point
separating delocalized and localized phases is
found.\cite{Kim_disorder,Kim_disorder_sigma} The RG flow shows
that the effects of random potentials vanish if we start from
sufficiently weak disorder. The stability of the charged critical
point against weak disorder in the QED$_3$, physically, results
from the fact that the fermionic spinons feel the effective
dimensionality higher than two owing to the long range gauge
interactions at the charged critical point, thus killing the
effects of weak disorder.\cite{Kim_disorder_sigma}

In the spinon-gauge critical theory Eq.
(26)\cite{Kim_Deconfine_FS} a similar result is expected.
Deconfined spinons near the Fermi surface would remain delocalized
at least against weak randomness owing to long range gauge
interactions while noninteracting spinons without gauge
interactions are localized by random potentials according to the
scaling theory.\cite{MAMIT} However, it should be considered that
the presence of nonmagnetic disorder reduces the spinon
conductivity $\sigma_{0}$. Thus, even if the charged fixed point
can be stable against weak disorder in the case of noncompact U(1)
gauge fields, the fixed point can be unstable against instanton
excitations owing to the reduction of the
conductivity.\cite{Kim_Deconfine_FS} As mentioned earlier, the
spin liquid phase can be stable when the spinon conductivity is
sufficiently large. In the spinon bad metal the spinons would be
confined owing to the presence of disorder. This is the main
reason why the application of the present slave-rotor formulation
should be limited within weak disorder.

The above discussion seems that the spin liquid Mott insulator
remains stable against weak randomness in contrast to the
emergence of the spin liquid Bose glass in the first
treatment.\cite{Boson_Fermion} To interpret this inconsistency
between the two approaches in a consistent manner, we claim that
the U(1) spin liquid of spinons is stable against weak randomness,
but the Bose-Mott insulator of chargons is not. The resulting
insulator is identified with the spin liquid charge glass.

\section{Summary}

In the present paper we examined the role of disorder in the
Mott-Hubbard metal-insulator transition based on the slave-rotor
formulation of the Hubbard model. In this representation the
Mott-Hubbard insulator is understood as the spin liquid Mott
insulator in terms of gapless spinons and gapped chargons
interacting via U(1) gauge fields. We found that the Mott-Hubbard
critical point becomes unstable as soon as disorder is turned on,
resulting in a disorder critical point interpreted as the spin
liquid glass insulator to the Fermi liquid metal transition. The
glassy behaviors of charge fluctuations\cite{Fisher_BG} can be
measured by the optical spectra in the insulating phase of an
organic material $\kappa-(BEDT-TTF)_{2}Cu_{2}(CN)_{3}$.
Furthermore, since the Mott-Anderson critical points depend on the
spinon conductivity, universality in the critical exponents may
not be found. Last, we open the possibility that the spin liquid
Mott insulator may survive against weak randomness.\cite{BM_BG}

\section*{Acknowledgement}

K. -S. Kim acknowledges that the condensed matter group meeting
motivates him to investigate this problem, and thanks the members,
Dr. G.-S. Jun, J.-H. Han, K. Park, and J.-W. Lee.


\begin{thebibliography}{9}
\bibitem{Derivation} One can rewrite Eq. (20) as \bqa &&
e_{vr}^{2} = e^{l}e^{lnZ_{c}}e_{v}^{2} , \nn && u_{vr} =
e^{l}e^{lnZ_{u}^{-1}}e^{lnZ_{\Phi}^{2}}u_{v} , \nn && W_{r} =
e^{2l}e^{lnZ_{W}^{-1}}e^{lnZ_{\Phi}^{2}}W . \nonumber \eqa
Differentiating the above scaling equations with respect to the
scaling parameter $l$, one obtains the exact scaling equations
\bqa && \frac{de_{vr}^{2}}{dl} = e_{vr}^{2} +
\frac{dlnZ_{c}}{dl}e_{vr}^{2} , \nn && \frac{du_{vr}}{dl} = u_{vr}
+ \frac{dlnZ_{u}^{-1}}{dl}u_{vr} +
\frac{dlnZ_{\Phi}^{2}}{dl}u_{vr} , \nn && \frac{dW_{r}}{dl} =
W_{r} + \frac{dlnZ_{W}^{-1}}{dl}W_{r} +
\frac{dlnZ_{\Phi}^{2}}{dl}W_{r} . \nonumber \eqa The
renormalization factors result from singular contributions of
dynamical processes, formally given in the one loop
approximation\cite{Weak_disorder_fixed_point,Herbut_BG,Kleinert,Ye}
\bqa && Z_{c} \approx exp\Bigl[-\Bigl({\lambda} +
\frac{\zeta}{\sigma_{0}}\Bigr)e_{vr}^{2}l\Bigr] , \nn && Z_{u}
\approx  \nn &&
exp\Bigl[-\Bigl([h(\sigma_0,e_{vr}^{2})-2c_{\Phi}]e_{vr}^{2} +
\gamma{W}_{r} - \rho{u}_{vr} -
g(\sigma_{0},e_{vr}^{2})\frac{e_{vr}^{4}}{u_{vr}} \Bigr)l\Bigr] ,
\nn && Z_{W} \approx
exp\Bigl[-\Bigl([h(\sigma_0,e_{vr}^{2})-2c_{\Phi}]e_{vr}^{2} -
\kappa{u_{vr}} + \eta{W}_{r}\Bigr)l\Bigr] , \nn && Z_{\Phi}
\approx exp\Bigl[c_{\Phi}e_{vr}^{2}l\Bigr] , \nonumber \eqa where
all numerical values are positive.
\bibitem{RG_fixed_point} The weak disorder fixed point is given by
$u_{v}^{*}= (\eta - 2\gamma)/(\eta\rho - \gamma\kappa)$ and $W^{*}
= (\kappa - 2\gamma)/(\eta\rho -\gamma\kappa)$. For this fixed
point to exist, the positive numerical constants should satisfy
$\eta\rho<\gamma\kappa<2\gamma\rho$ or
$\eta\rho>\gamma\kappa>2\gamma\rho$. Using the numerical values in
Refs. \cite{Weak_disorder_fixed_point,Herbut_BG}, one can check
that $\eta\rho<\gamma\kappa<2\gamma\rho$ is satisfied.
\bibitem{RG_fixed_point2} The weak disorder fixed point is given by
$e_{v}^{2*} = \sigma_{0}/(\lambda\sigma_{0}+\zeta)$, $u_{v}^{*}=
(\eta - 2\gamma+(\eta-\gamma)c_{h}\sigma_{0}) /(\eta\rho -
\gamma\kappa)$ and $W^{*} = (\kappa - 2\gamma +
(\kappa-\gamma)c_{h}\sigma_{0})/(\eta\rho -\gamma\kappa)$, where
$\eta\rho<\gamma\kappa<\gamma\rho$ is satisfied.
\bibitem{Boson_Fermion} This difference between the fermion-gauge
and boson-gauge systems for the role of disorder originates from
the fact that the scaling dimension of boson fields differs from
that of fermion fields. As mentioned before, the quadratic
derivative in the boson action yields the bare scaling dimension
of $L^{1/2}$ to the boson field while the linear derivative in the
fermion action, resulting from the expansion near the Fermi
surface, gives $L$ to the fermion field. This difference causes
that the bare scaling dimension of disorder is $L^{2}$ in the
boson system while it is $L^{0}$ in the fermion system. Thus, the
charged critical point in the boson-gauge action becomes unstable
as soon as disorder is turned on, as discussed in the text. On the
other hand, the charged fixed point in the fermion system remains
stable against weak disorder because gauge interactions make the
scaling of disorder weakly negative.\cite{Kim_disorder_sigma} This
is the statement that fermions feel the effective dimensionality
higher than two, thus killing the effects of weak disorder. This
concept is analogous to the Harris criterion determining the
stability of the critical point against weak randomness.
\end{thebibliography}
\end{document}